\documentclass[10pt]{article}{\textwidth=125mm}{\textheight=195mm}
\usepackage{amsmath,amssymb,amsthm,pb-diagram,lamsarrow,pb-lams,  hyperref}
\usepackage{graphicx}
\usepackage{tabularx}
\usepackage{float}
\usepackage{listings}

\title{Top-down Paradigm in Engineering Software Integration}

\theoremstyle{plain}

\chardef\bslash=`\\ 

\theoremstyle{definition}

\theoremstyle{remark}

\newbox\ncintdbox \newbox\ncinttbox 
\setbox0=\hbox{$-$} \setbox2=\hbox{$\displaystyle\int$}
\setbox\ncintdbox=\hbox{\rlap{\hbox
    to \wd2{\hskip-.125em \box2\relax\hfil}}\box0\kern.1em}
\setbox0=\hbox{$\vcenter{\hrule width 4pt}$}
\setbox2=\hbox{$\textstyle\int$} \setbox\ncinttbox=\hbox{\rlap{\hbox
    to \wd2{\hskip-.175em \box2\relax\hfil}}\box0\kern.1em}

\begin{document}
\maketitle \setlength{\parindent}{0pt}
\begin{center}
\author{}
{\textbf{Petr R. Ivankov*}\\
e-mail: * monstr3d@korolev-net.ru \\
}
\end{center}

\vspace{1 in}

\begin{abstract}
\noindent

The top-down approach of engineering software integration is considered in this parer. A set of advantages of this approach are presented, by examples. All examples are supplied by open source code.

\end{abstract}

\

\section{Introduction}

\begin{figure}[htp]
\begin{center}
\hspace{-0.2cm}
\includegraphics[scale=0.25]{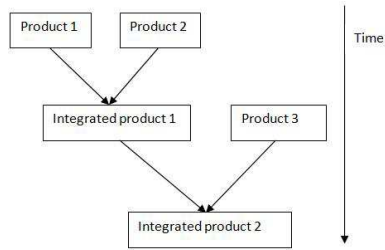}
\caption{Chaotic integration of software products}\label{fig:chaoticinteg}
\end{center}
\end{figure}

Historically engineering software integration is rather chaotic then planned. Different CAD products had been integrated with CAE and CAM products, Matlab had been integrated with Simulink etc. This integration history could be schematically presented in figure \ref{fig:chaoticinteg}.

\begin{figure}[htp]
\begin{center}
\hspace{-0.2cm}
\includegraphics[scale=0.7]{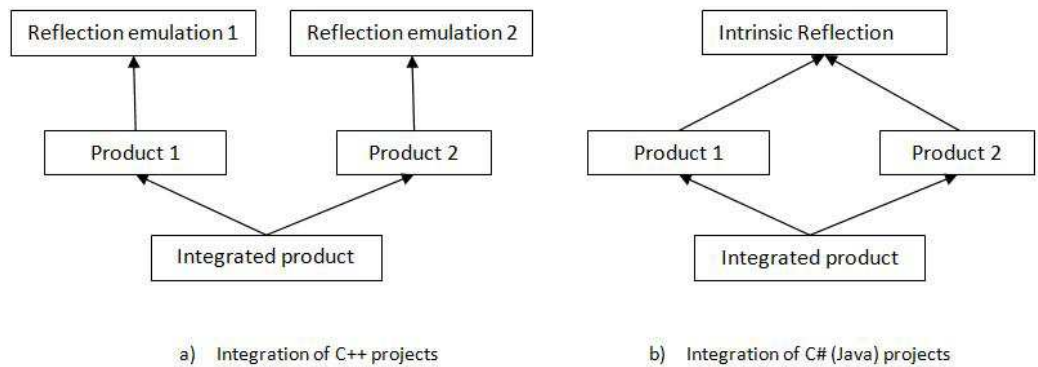}
\caption{Reflection emulation and intrinsic reflection}\label{fig:reflection_implementation}
\end{center}
\end{figure}

However chaotic way is not optimal way. For example C++ programming language had been designed chaotically. Languages of next generation (Java, C\#) had been designed by planned way. C++ does not support automatic garbage collection. But any present day project should have it. So big C++ projects contain smart pointers those provide automatic garbage collection. Since smart pointer is not intrinsic C++ feature there exists a lot of versions of smart pointers. Integration of set of projets with different versions of smart pointers has a set of disadvantages. In this case integrated product is not clear, volume of its code is overloaded etc. Present day programming languages (Java, C\#) support Reflection but C++ does not support it. However Reflection is very useful feature for any big project. So big C++ projects have different versions of Reflection emulation. Comparting of intrinsic reflection and reflection emulation is presented in figure \ref{fig:reflection_implementation}.

\begin{figure}[htp]
\begin{center}
\hspace{-0.2cm}
\includegraphics[scale=0.7]{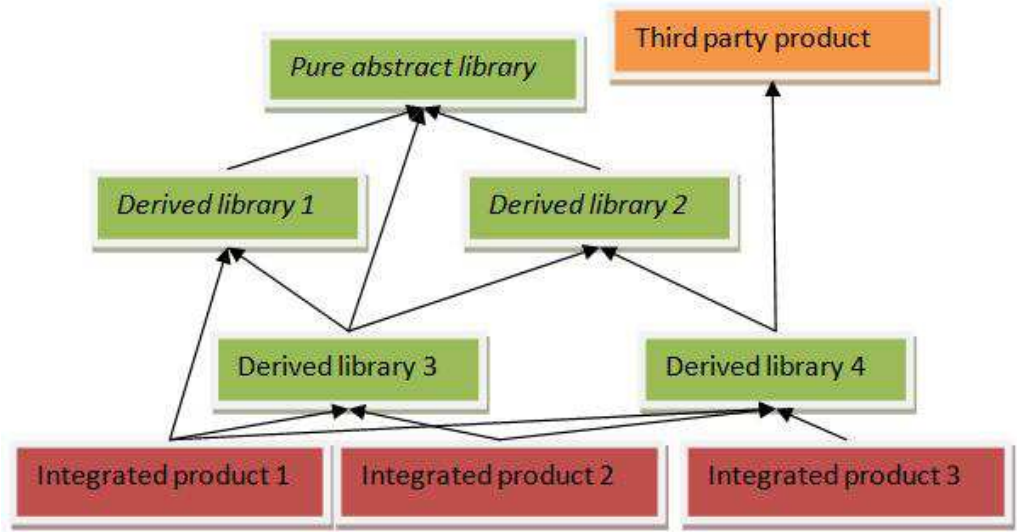}
\caption{Top-down integration of engineering software products}\label{fig:topdowninteg}
\end{center}
\end{figure}

In case of emulation we have additional code without essentially new functionality. If we would like integrate a set of projects by such way then we will obtain a lot of lumber. Such disadvantage have any chaotic way of integration. But chaotic way has top-down design alternative. The top-down paradigm is presented in figure \ref{fig:topdowninteg}.

Let us explain meaning of figure \ref{fig:topdowninteg}. First of all we have pure abstract library devoted to science and engineering. This library is used by derived libraries devoted to different brunches of science and engineering. Libraries can be abstract. Integration can use other engineering software as third party. Libraries are being used by integrated products. Green, pig and purple color are used for libraries, third party and integrated products respectively. Italic font is used for abstract libraries.

\section{Interoperability issues}
\begin{figure}[htp]
\begin{center}
\hspace{-0.2cm}
\includegraphics[scale=0.4]{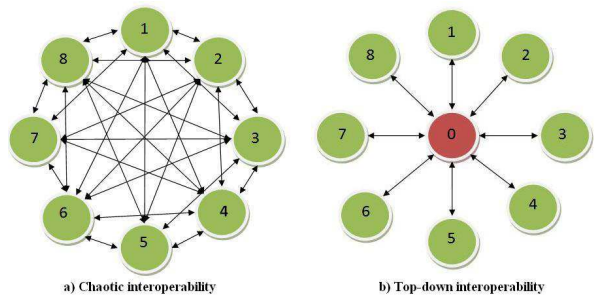}
\caption{Interoperability}\label{fig:interoperability_issues}
\end{center}
\end{figure}

Chaotic way interoperability disadvantage is well known. Chaotic way of interoperability of $N$ products requires development of $\frac{N(N-1)}{2}$ adapters. Adapters provide interoperability between objects of any pair of $N$ element set. In figure \ref{fig:interoperability_issues} (a) we have 8 products. Two sided arrows are adapters.
If we use top-down scheme then we should develop adapters between base object and other objects. So we should have $N$ adapters only. The base object is 0 - object in figure   \ref{fig:interoperability_issues} (b).
\section{Category theory as prototype}
\begin{figure}[htp]
\begin{center}
\hspace{-0.2cm}
\includegraphics[scale=0.07]{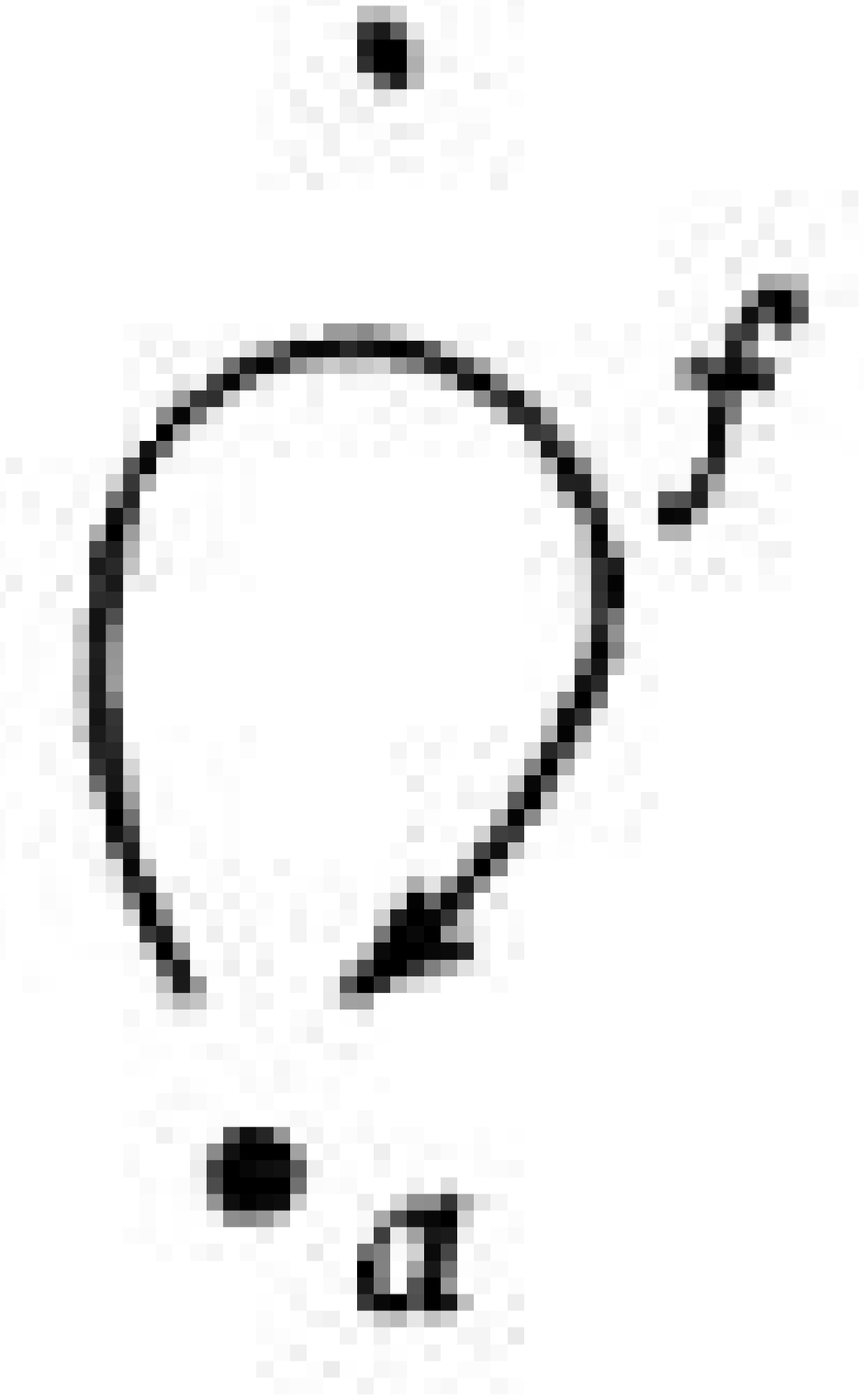}
\caption{Object and arrow}\label{fig:object_and_arrow}
\end{center}
\end{figure}

Pure abstract library for science and engineering should have very high level of abstraction. However this level of of abstraction is already developed in math. In mathematics, category theory deals in an abstract way with mathematical structures and relationships between them: it abstracts from sets and functions to objects linked in diagrams by morphisms or arrows.  High level of abstraction of Category Theory is explained in \cite{topoi}. This book contains figure \ref{fig:object_and_arrow} with following text.
\footnotesize

\begin{figure}[htp]
\begin{center}
\hspace{-0.2cm}
\includegraphics[scale=0.4]{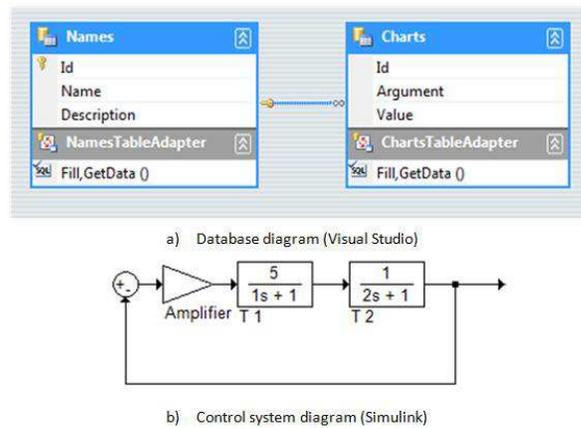}
\caption{Database and control system diagrams}\label{fig:intrinsic_database_controlsystems}
\end{center}
\end{figure}

We did not actually say what $a$ and $f$ are. The point is that they can be anything you like. $a$ might be a set with $f$ its identity function. But $f$ might be a number, or a pair of numbers, or a banana, or the Eiffel tower, or even Richard Nixon. Likewise for $a$. 
\newline
\normalsize
Let us consider application of this abstraction. Database diagram and control systems diagram are presented in figure \ref{fig:intrinsic_database_controlsystems}

\begin{figure}[htp]
\begin{center}
\hspace{-0.2cm}
\includegraphics[scale=0.4]{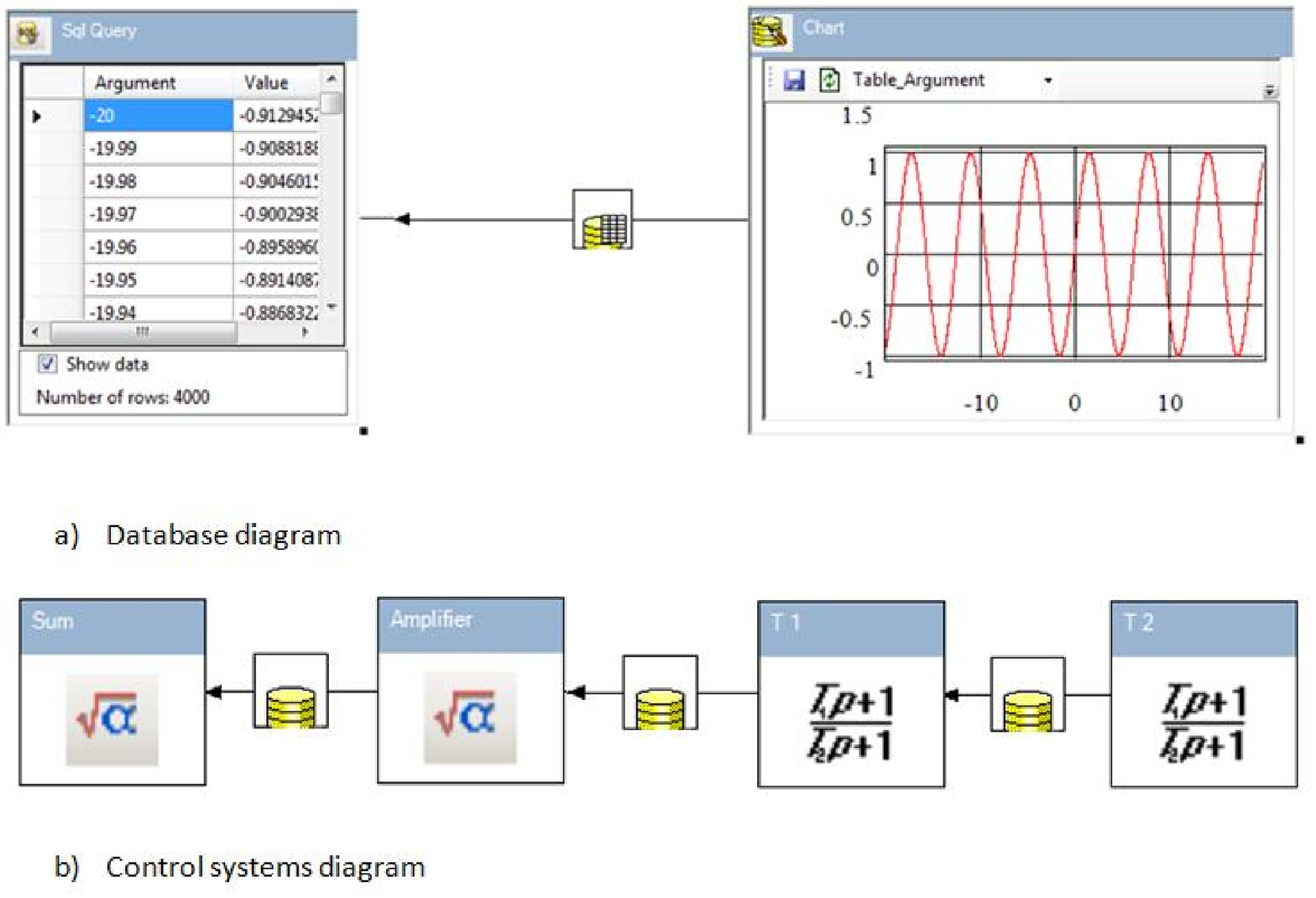}
\caption{Database and control system diagrams in single framework}\label{fig:common_database_controlsystems}
\end{center}
\end{figure}

Common features of these diagram are objects and arrows as abstract objects. Figure \ref{fig:common_database_controlsystems} presents database and control systems diagram in single framework (There is no one to one map).

\begin{figure}[htp]
\begin{center}
\hspace{-0.2cm}
\includegraphics[scale=0.4]{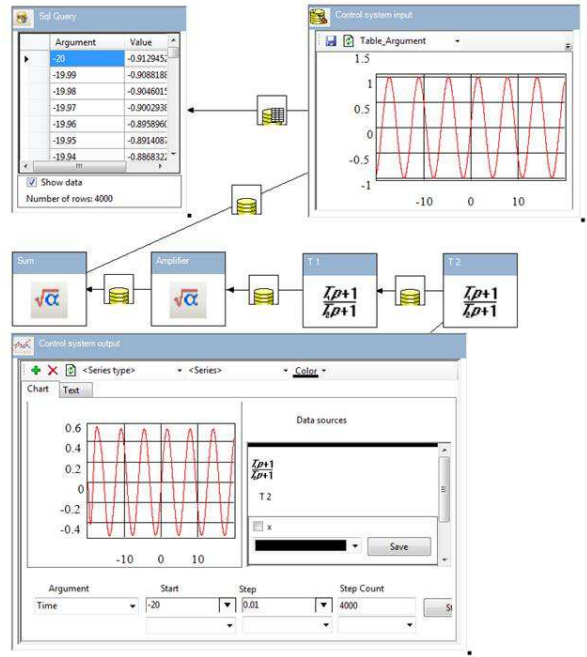}
\caption{Integrated database and control system diagram}\label{fig:integrated_database_controlsystems}
\end{center}
\end{figure}

Now we would like to integrate both diagrams. Integrated diagram is presented in figure \ref{fig:integrated_database_controlsystems}
Presented on \ref{fig:integrated_database_controlsystems} diagram has following meaning. Database contains recorded input signals of control systems. This diagram enable us to define response of control system.

\section{Present day state}
\begin{figure}[htp]
\begin{center}
\hspace{-0.2cm}
\includegraphics[scale=0.5]{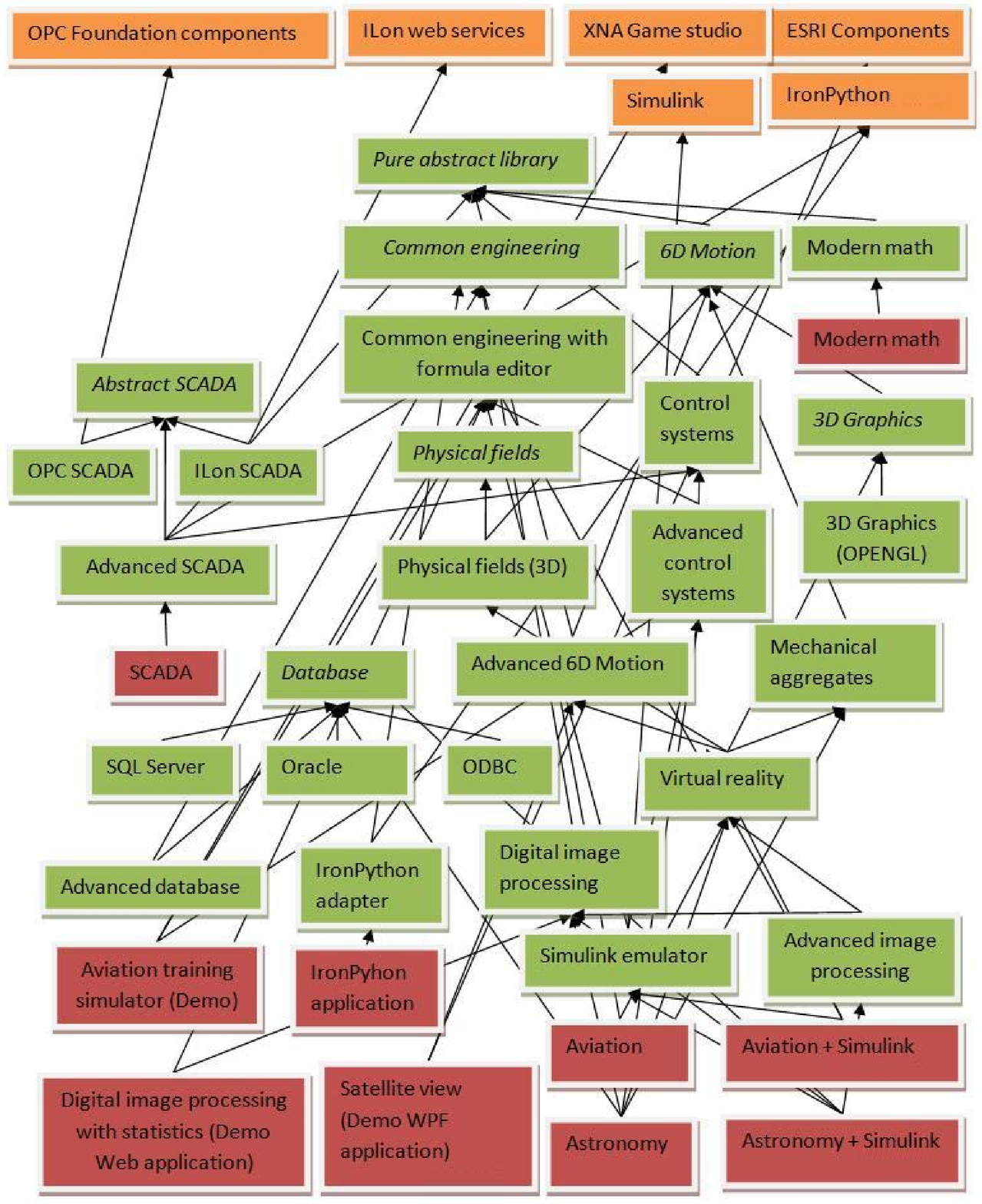}
\caption{Hierarchy of assemblies}\label{fig:product_hierarchy}
\end{center}
\end{figure}

Explanation of idea usefulness is difficult without any implementation. This idea shall find further development. However a lot of features are already implemented. These features are presented in figure \ref{fig:product_hierarchy}.
Presented in figure \ref{fig:product_hierarchy} scheme reflects only part of implemented features. Source code can be downloaded from \url{http://www.mathframe.com}. It is worth to note whole product is not a sum of presented in figure \ref{fig:product_hierarchy} features only.  These features can interact and interaction make product much more powerful. This thesis will be shown below by examples. This hierarchy can inspire some questions. One of reasonable question is: "Why we need abstract physical field library?" Now this software supports physical fields in 3D space. However this software is declared as universal and should support 2D physical fields in future. And 2D fields library shall be inherited from abstract physical field library. Universality had been provided from the very beginning of development. Demo applications show prospects of further development.
\section{Elementary examples}
Examples of this section are not not related to real science and engineering problems. These examples are rather textbook which is very facile for idea explanation.
\subsection{Physical fields}
\begin{figure}[htp]
\begin{center}
\hspace{-0.2cm}
\includegraphics[scale=0.2]{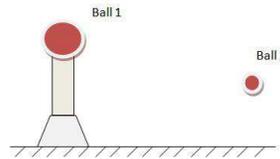}
\caption{Charged balls}\label{fig:charged_balls}
\end{center}
\end{figure}

\begin{figure}[htp]
\begin{center}
\hspace{-0.2cm}
\includegraphics[scale=0.5]{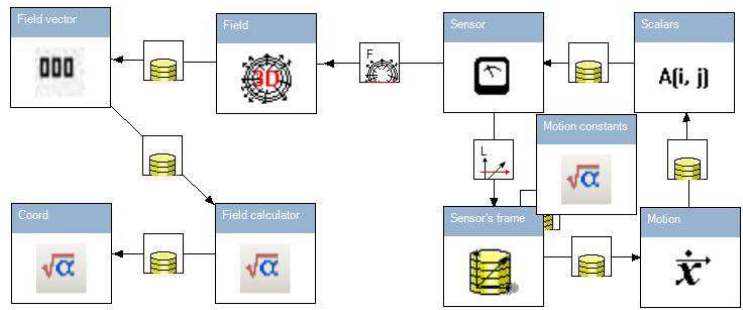}
\caption{Charged balls motion simulation}\label{fig:charged_balls_calculation}
\end{center}
\end{figure}

\begin{figure}[htp]
\begin{center}
\hspace{-0.2cm}
\includegraphics[scale=0.4]{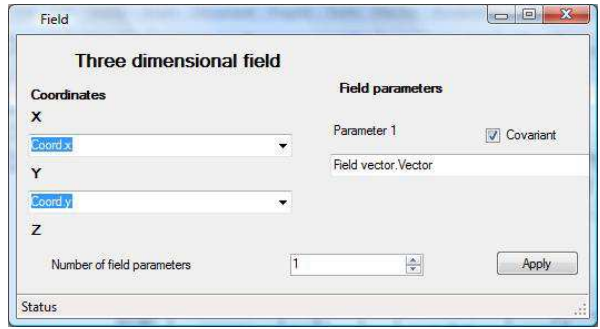}
\caption{Properties of $\mathbf{Field}$ object}\label{fig:field_object}
\end{center}
\end{figure}

One physical phenomenon acts to another one and there is backward dependence. This thesis can be exhibited by example presented in figure \ref{fig:charged_balls}.
We have two charged balls $\mathbf{Ball\ 1}$ and $\mathbf{Ball\ 2}$. Electrostatic force which act to $\mathbf{Ball\ 2}$ depends on relative position $\mathbf{Ball\ 2}$ with respect to $\mathbf{Ball\ 1}$. Otherwise the relative position as time function depend on motion of $\mathbf{Ball\ 2}$ and therefore relative position depends on electrostatic force. Common consideration of 6D motion and physical fields make software much more effective. Now we would like simulate motion of $\mathbf{Ball\ 2}$. Simulation scheme is presented in figure \ref{fig:charged_balls_calculation}.
Left part of this picture represents electrostatic field of $\mathbf{Ball\ 1}$. Objects $\mathbf{Coord}$, $\mathbf{Field\ calculator}$ and $\mathbf{Field\ vector}$ provide calculations of next formula of electrostatic intensity:
\begin{equation}\nonumber
E = \frac{kr}{|r|^3};
\end{equation}
where $r$ is 3D vector of relative position with respect to charge, $k$ is coefficient which depends on charge of $\mathbf{Ball\ 1}$ and system of units.
This calculation is separated for performance issue. Object $\mathbf{Coord}$ calculates $\frac{k}{|r|^3}$, object  $\mathbf{Field\ calculator}$ calculates components of 3D vector $\frac{kr}{|r|^3}$ and $\mathbf{Field\ vector}$ assemblies these components into vector. All these calculation are used by $\mathbf{Field}$ object. Properties of the  $\mathbf{Field}$  objet are presented on figure \ref{fig:field_object}.

\begin{figure}[htp]
\begin{center}
\hspace{-0.2cm}
\includegraphics[scale=0.4]{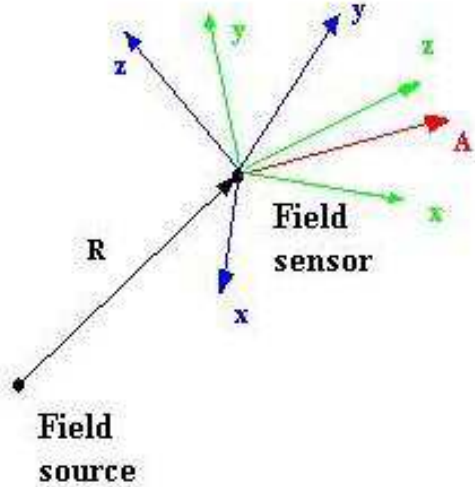}
\caption{Covariant field}\label{fig:covariant_field}
\end{center}
\end{figure}

\begin{figure}[htp]
\begin{center}
\hspace{-0.2cm}
\includegraphics[scale=0.3]{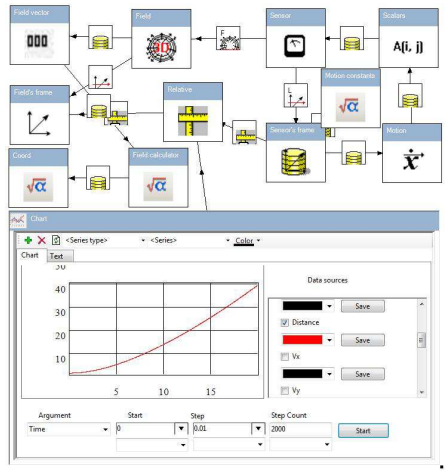}
\caption{Charged balls motion simulation with relative measurements}\label{fig:charged_balls_calculation_relative}
\end{center}
\end{figure}

\begin{figure}[htp]
\begin{center}
\hspace{-0.2cm}
\includegraphics[scale=0.2]{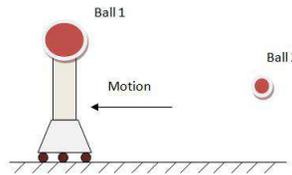}
\caption{Charged balls with moved ball 1}\label{fig:charged_balls_motion}
\end{center}
\end{figure}

These properties have following meaning. First of all field parameter is output vector of  $\mathbf{Field\ vector}$. Figure \ref{fig:covariant_field} explains meaning of "covariant" term.
If 3D vector is not covariant then its components depend on sensor position only. Covariant vector components depend on both orientation and position. Values of components are projections of geometric vector to sensor's axes of reference. Figure \ref{fig:covariant_field} presents two orientations of sensor: blue and green. Projections of field vector $\mathbf{A}$ are different for these different orientations. So we have 3D covariant physical field. The $\mathbf{Sensor}$ object is a sensor of the field. The $\mathbf{Scalars}$ object provides components of field vector obtained by the $\mathbf{Sensor}$ object. These components are used in following motion equations of $\mathbf{Ball\ 2}$:
\begin{equation}\nonumber
\dot{x}=V_x;
\end{equation}
\begin{equation}\nonumber
\dot{y}=V_y;
\end{equation}
\begin{equation}\nonumber
\dot{z}=V_z;
\end{equation}
\begin{equation}\nonumber
\dot{V}_x=aE_x;
\end{equation}
\begin{equation}\nonumber
\dot{V}_y=aE_y;
\end{equation}
\begin{equation}\nonumber
\dot{V}_z=aE_z.
\end{equation}
where $x$, $y$, $z$ are coordinates of $\mathbf{Ball\ 2}$ and $V_x$, $V_y$, $V_z$ are components of its velocity. Coefficient $a$ depends of charge of  $\mathbf{Ball\ 2}$ and system of units. These motion equations are contained in $\mathbf{Motion}$ object. Then  coordinates $x$, $y$, $z$ are used by moved reference frame $\mathbf{Sensor's\ frame}$. The $\mathbf{Sensor}$ object is installed on $\mathbf{Sensor's\ frame}$. So we have backward dependence. Motion and therefore current position of $\mathbf{Sensor's\ frame}$ and $\mathbf{Sensor}$ depends on current field value. Otherwise current field value depends on current position. Let us  make situation more complicated. New situation is presented on figure \ref{fig:charged_balls_calculation_relative}
Here we have installed $\mathbf{Field}$ on $\mathbf{Field's\ frame}$ and added new object $\mathbf{Relative}$. The $\mathbf{Relative}$ defines relative motion parameters of $\mathbf{Sensor's\ frame}$ with respect to $\mathbf{Field's\ frame}$. Relative distance is indicated by $\mathbf{Chart}$ object. Physical picture remains the same. We have added indication only. Now let us install $\mathbf{Ball\ 1}$ on moved platform as it is presented in figure \ref{fig:charged_balls_motion}

\begin{figure}[htp]
\begin{center}
\hspace{-0.2cm}
\includegraphics[scale=0.5]{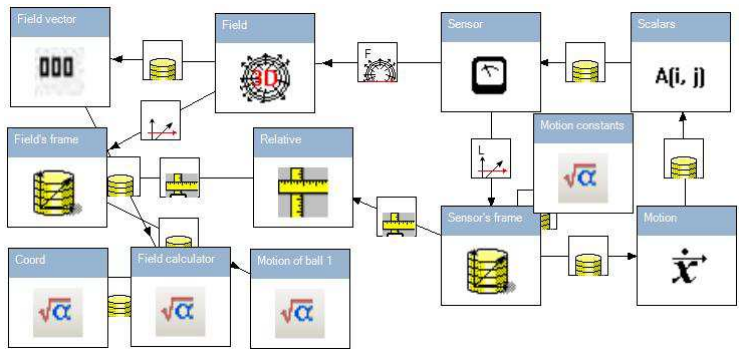}
\caption{Charged balls motion simulation with moved ball 1}\label{charged_balls_motion_platform}
\end{center}
\end{figure}

Simulation of this situation is presented on figure \ref{charged_balls_motion_platform}.

Now $\mathbf{Field's\ frame}$ is moved and $\mathbf{Motion\ of\ ball\ 1}$ object defines motion low. The time dependence of distance between $\mathbf{Ball\ 1}$ and $\mathbf{Ball\ 2}$ has been changed by following factors. First of all now  $\mathbf{Ball\ 1}$ which is source of $\mathbf{Field}$ is moved. Its own motion influence on relative distance. Secondly motion of field source influence on field intensity near $\mathbf{Ball\ 2}$. So absolute motion of $\mathbf{Ball\ 2}$ will be changed. This factor also influence on the distance. These facts are physically evident. I would like to exhibit facilities of such approach from point of view of software user. The user simply has added motion to $\mathbf{Ball\ 1}$ and all other dependent factors had been automatically taken into account. This facility is impossible without integration.
\subsection{Space aerodynamics and digital image processing}
\begin{figure}[htp]
\begin{center}
\hspace{-0.2cm}
\includegraphics[scale=0.2]{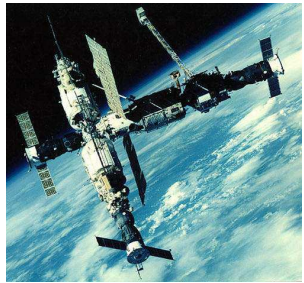}
\caption{The Mir orbital station}\label{fig:orbital_station}
\end{center}
\end{figure}

\begin{figure}[htp]
\begin{center}
\hspace{-0.2cm}
\includegraphics[scale=0.2]{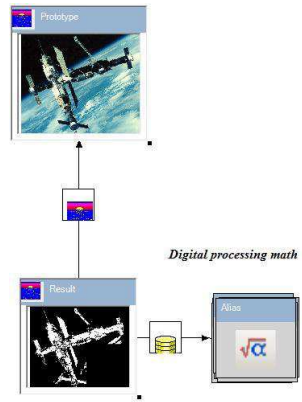}
\caption{The Mir orbital station digital image processing}\label{fig:orbital_station_digital_processing}
\end{center}
\end{figure}

The key feature of space aerodynamics is that spacecraft interacts with molecules which do not collide with each other. Therefore aerodynamic force depends on visible area and does not depend on other parameters of spacecraft shape. We can use digital image processing for space aerodynamic calculation.  Let us consider it for space aerodynamics. The image of the Mir orbital station is presented on figure  \ref{fig:orbital_station}.
This photo should be filtered for space aerodynamics usage. This filtering (digital image processing) is presented on figure \ref{fig:orbital_station_digital_processing}.
The $\mathbf{Prototype}$ object contains source image and $\mathbf{Result}$ contains filtering result. Other objects contain necessary math. This sample presents main advantage of integration. Software devoted to space technology only is not effective without digital image processing. Otherwise digital image processing is not effective without advanced math.
\subsection{Algebraic topology}
\begin{figure}[htp]
\begin{center}
\hspace{-0.2cm}
\includegraphics[scale=0.2]{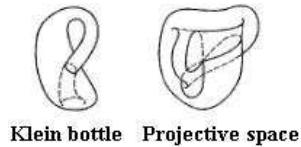}
\caption{Klein bottle and projective space}\label{fig:klein_bottle_and_projective_space}
\end{center}
\end{figure}

\begin{figure}[htp]
\begin{center}
\hspace{-0.2cm}
\includegraphics[scale=0.2]{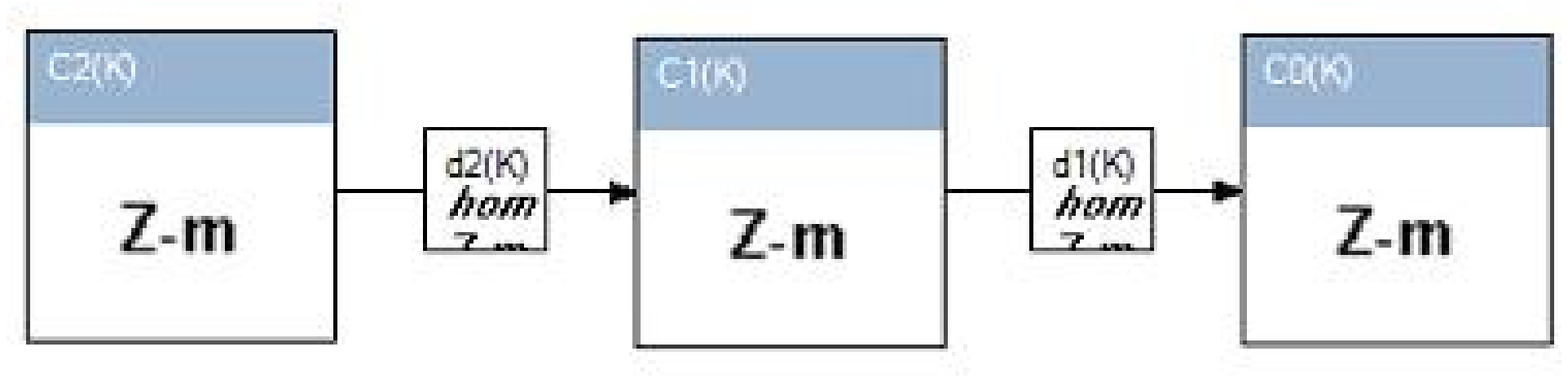}
\caption{Klein bottle chain complex representation}\label{fig:klein_bottle_chains}
\end{center}
\end{figure}

Algebraic topology is not yet everyday tool of engineer or scientist. Now this branch of science looks rather as exotic. However exotic tasks are good tests for integrability and universality. The "Modern math" has been developed as test of prospects of Category Theory approach. In this chapter calculation of different topological invariants \cite{spanier} is considered. Let us consider Klein bottle $K$ and real projective space $\mathbb{R}P^3$ (figure \ref{fig:klein_bottle_and_projective_space}
Homology groups of these spaces can be calculated as homology of following chain complexes \cite{spanier}:
\[
\begin{diagram}
    \node{...} \arrow{e,t}{\partial_3}\node{C_2} \arrow{e,t}{\partial_2}  \node{C_1} \arrow{e,t}{\partial_1} \node{C_0;}
\end{diagram}
\]
Chain complex of Klein bottle $K$ and real projective space is presented below:
\[
\begin{diagram}
    \node{\mathbb{Z}} \arrow{e,t}{0}\node{\mathbb{Z}} \arrow{e,t}{2}  \node{\mathbb{Z}} \arrow{e,t}{0} \node{\mathbb{Z}}
\end{diagram}
\]
The application representation of it is shown in figure \ref{fig:klein_bottle_chains}

Chain complex of real projective space $\mathbb{R}P^3$ is presented below:
\[
\begin{diagram}
    \node{\mathbb{Z}} \arrow{e,t}{(\begin{smallmatrix}1\\1\end{smallmatrix})}\node{\mathbb{Z}\times \mathbb{Z}} \arrow{e,t}{(1,1)} \node{\mathbb{Z}}
\end{diagram}
\]

\begin{figure}[htp]
\begin{center}
\hspace{-0.2cm}
\includegraphics[scale=0.2]{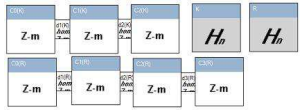}
\caption{Klein bottle chain complex representation}\label{fig:k_and_r_chains}
\end{center}
\end{figure}

Chain complexes of both spaces are presented in figure \ref{fig:k_and_r_chains}.

Now we can calculate homology and cohomology groups, homology with coefficients and other invariants. More details you can find at Category Theory project \url{http://categorytheory.sourceforge.net/}.
\section{Realistic samples}
Here we consider more complicated samples which are related to real engineering problems. This problems include following disciplines:
\newline
- advanced mechanics;
\newline
- processing of signals;
\newline
- statistics;
\newline
- system identification;
\newline
- control theory;
\newline
- celestial navigation;
\newline
- astronomy;
\newline
- geomagnetism;
\newline
- space aerodynamics;
\newline
- digital image processing;
\newline
- virtual reality.
All these samples can be downloaded from \url{http://www.codeproject.com/KB/architecture/grandiose2.aspx}
\subsection{Advanced mechanics}\label{sec:adv_mech}
\begin{figure}[htp]
\begin{center}
\hspace{-0.2cm}
\includegraphics[scale=0.2]{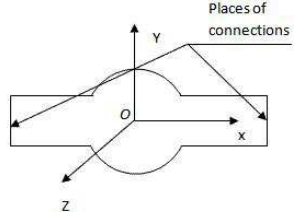}
\caption{Spacecraft module}\label{fig:spacecraft_module}
\end{center}
\end{figure}

\begin{figure}[htp]
\begin{center}
\hspace{-0.2cm}
\includegraphics[scale=0.3]{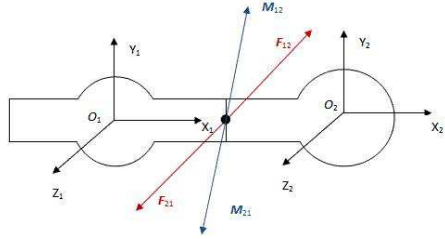}
\caption{Connection of two modules}\label{fig:spacecraft_aggregate}
\end{center}
\end{figure}

Space technology provides good samples of advanced mechanics. Orbital station (figure \ref{fig:orbital_station}) is a very complicated mechanical object. It is not rigid body. It has solar cell panels. These panels are elastic. The station is stabilized by gyros. Moreover station configuration is not constant. The "Mechanical aggregate" library had been developed for simulation of similar aggregates. The library contains aggregate designer. Why aggregate designer? Indeed mechanical equations are well known long time ago. But software development for simulation of complicated mechanical objects is not quite easy task. Aggregate designer make this task much easier. Aggregate designer is integrated into framework. This fact enables us provide interpretability of mechanics with physical fields. So it is easy to simulate action of magnetic fields on mechanical objects. I will consider this task below. Now we would like to create mechanical model of spacecraft from models of its modules. Typical spacecraft module is schematically presented in figure \ref{fig:spacecraft_module}
This module has own coordinates system $OXYZ$. Also it has places of connections. We can connect other modules to this module. Behavior of module is defined by following kinematic parameters:
\newline
- radius vector $r$;
\newline
- velocity $V$;
\newline
- orientation quaternion $Q$;
\newline
- angular velocity $\omega$.
\newline
But module is not rigid in general. And these parameters are not parameters of module. These parameters are rather parameters of one point of module. In this article we suppose that these parameters are parameters of origin of $OXYZ$ coordinates system. Since module is not rigid it has additional degrees of freedom. These degrees of freedom can be interpreted as generalized coordinates $q_i$ ($i$= 1,...,$n$). Instant state of module is defined by following parameters: $r$, $Q$, $q_1$, ..., $q_n$, $V$, $\omega$, $\dot{q}_1$, ..., $\dot{q}_n$. This parameters will be named state variables. Parameters $\dot{V}$, $\dot{\omega}$, $\ddot{q}_1$, ..., $\ddot{q}_n$ will be called accelerations. Mechanical equations define accelerations by state parameters. Accelerations near connection can be defined by following way:
\begin{equation}\label{v_eq}
\dot{\omega_i}=\varepsilon_i + P_{\omega_i}\dot{V}+Q_{\omega_i}\dot{\omega}+\sum_k R_{\omega_{ik}}\ddot{q}_k;
\end{equation}
\begin{equation}\label{o_eq}
\dot{V_i}=a_i + P_{V_i}\dot{V}+Q_{V_i}\dot{\omega}+\sum_k R_{V_{ik}}\ddot{q}_k.
\end{equation}
where $i$ is number of connection. Other variables are matrixes which depend on state variables. Connection of two modules is presented on figure \ref{fig:spacecraft_aggregate}.
Both modules have equal acceleration near connection. First module acts to second one by force $F_{12}$ and mechanical momentum $M_{12}$. Similarly second module acts to first one by force $F_{21}$ and mechanical momentum $M_{21}$. These forces and momentums satisfy following Newton equations:
\begin{equation}\label{F_eq}
F_{21}=-F_{21};
\end{equation}
\begin{equation}\label{M_eq}
M_{21}=-M_{21}.
\end{equation}
Mechanical equations of module can be represented by the following way:
\begin{equation}\label{VF_eq}
\dot{V}=a+\sum_i D_{V_i}F_i + \sum_i E_{V_i}M_i;
\end{equation}
\begin{equation}\label{oF_eq}
\dot{\omega}=a+\sum_i D_{\omega_i}F_i + \sum_i E_{\omega_i}M_i;
\end{equation}
\begin{equation}\label{qF_eq}
\ddot{q}_k=a+\sum_i K_{ki}F_i + \sum_i L_{ki}M_i.
\end{equation}
In these expressions accelerations are independent variables. $F_i$ ($M_i$) is force (mechanical momentum) of $i$- h connected module. Other vector and matrix parameters depend on state variables. Suppose that $m$ - h connection place $i$-h module is connected to $n$ - h connection place $j$-h module. Then we have following evident relations:
\begin{equation}\label{c_V}
\dot{V}_{im}=\dot{V}_{jn};
\end{equation}
\begin{equation}\label{c_o}
\dot{\omega}_{im}=\dot{\omega}_{jn};
\end{equation}
\begin{equation}\label{c_F}
F_{im}=-F_{jn};
\end{equation}
\begin{equation}\label{c_M}
M_{im}=-M_{jn}.
\end{equation}

\begin{figure}[htp]
\begin{center}
\hspace{-0.2cm}
\includegraphics[scale=0.3]{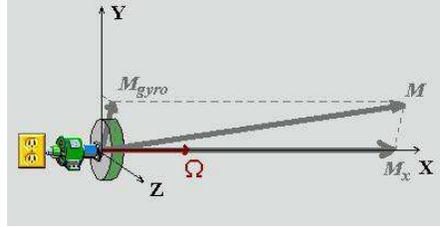}
\caption{Flywheel}\label{fig:flywheel}
\end{center}
\end{figure}

Expressions (\ref{v_eq}) - (\ref{c_M}) are  linear by accelerations system of equations. This system is fully defined and therefore can be solved. In result we have mechanical equations of whole aggregate. It is worth to note that this system is redundant. If $n_1$ ($n_2$) is number of freedom degrees of first (second) module then system has $n_1+n_2$ degrees of freedom. However aggregate has $n_1+n_2-6$ degrees of freedom. The exist software version that avoid this redundancy. But I will not describe it in this article. Aggregate designer provides equations of full aggregate automatically. Coefficients of (\ref{v_eq}) - (\ref{c_M}) equations are inputs of aggregate designer. In general these coefficients can depend on states of modules. There exist a lot of variants. But we can abstractly define calculation of these coefficients and then provide different implementations. Following types of mechanical modules are implemented:
\newline
- absolutely rigid body;
\newline
- elastic console;
\newline
- flywheel;
\newline
Equations of rigid body are well known and are not present them here. In context of aggregate designer rigid body has following properties:
\newline
- mass $m$;
\newline
- momentum of inertia $J$;
\newline
- number of connections;
\newline
\newline
- positions and orientations of connections.
\newline
These properties can be edited. Elastic console body is a mechanical system of infinite degrees of freedom. Usually math model of this object contains finite degrees of freedom with finite set of valuable harmonic oscillations \ref{harm_osc}. Every harmonic oscillation can be described by following second order system of ordinary differential equation:
\begin{equation}\nonumber
A\ddot{q}+\varepsilon\dot{q}+cq=Q.
\end{equation}
Where $Q$ is generalized force, $A$, $\varepsilon$ and $c$ are coefficients. Number of harmonics and their properties can be edited.
\newline
Flywheels are used in spin stabilization systems of spacecrafts. Following documents contain information devoted to spin stabilization systems:
\newline
- \url{http://www.freepatentsonline.com/3767139.html}
\newline
- \url{http://www.aiaa.org/content.cfm?pageid=406&gTable=mtgpaper&gID=58741}
\newline
- \url{http://adsabs.harvard.edu/abs/1966CosRe...4..173A}
\newline
\newline
Flywheel is forced by reversible engine (figure \ref{fig:flywheel}).

\begin{figure}[htp]
\begin{center}
\hspace{-0.2cm}
\includegraphics[scale=0.3]{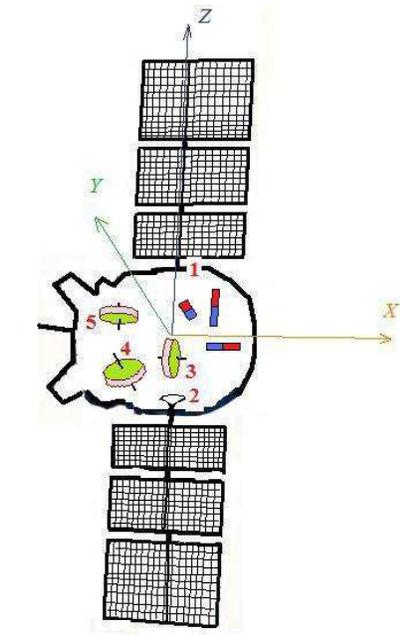}
\caption{Spacecraft with two elastic consoles and three flywheels}\label{fig:spacecraft_2_elastic_3_gyro}
\end{center}
\end{figure}

\begin{figure}[htp]
\begin{center}
\hspace{-0.2cm}
\includegraphics[scale=0.3]{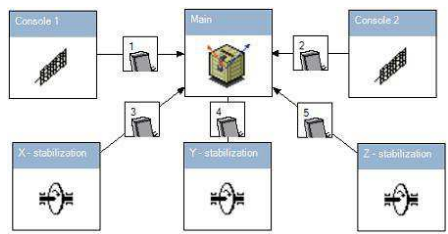}
\caption{Simulation model of spacecraft with two elastic consoles and three flywheels}\label{fig:spacecraft_2_elastic_3_gyro_simulation}
\end{center}
\end{figure}

Otherwise flywheel acts to engine by momentum $M_x$ . Since engine is attached to spacecraft this momentum is transferred to spacecraft. This momentum is used for spacecraft stabilization. Sine flywheel is rotated we have additional gyro momentum $M_{gyro}$ . Gyro momentum can be calculated by following expression:
\begin{equation}\nonumber
M_{gyro}=J_F\Omega\times\omega.
\end{equation}
where $J_F$  is inertial momentum of flywheel, $\Omega$ is angular velocity of flywheel and $\omega$ is angular velocity of engine (and also spacecraft). Total momentum $M$ is equal to geometric sum $M = M_x  + M_{gyro}$ ; Gyro momentum is undesirable factor. Stabilization system should require following condition $|M_{gyro}| << |M_x|$. However since engine acts to flywheel value of  is being increased by the time. Increasing of $M_{gyro}$ compensated by other devices which acts to spacecraft. In this article electromagnetic devices will be considered.
\newline
Now we can assembly simulation model of spacecraft presented on figure \ref{fig:spacecraft_2_elastic_3_gyro}.
Numbers 1 - 5 are numbers of connections places of spacecraft. Flywheels attached to 3, 4, 5 connection places realize angular stabilization of spacecraft with respect to axes $X$, $Y$, $Z$. Besides flywheels spacecraft contains three electromagnets for stabilization. Mechanical model of spacecraft is presented on figure \ref{fig:spacecraft_2_elastic_3_gyro_simulation}.
Numbers 1-5 of links are numbers of spacecraft connections.
\subsection{Virtual vibration test}
\begin{figure}[htp]
\begin{center}
\hspace{-0.2cm}
\includegraphics[scale=0.1]{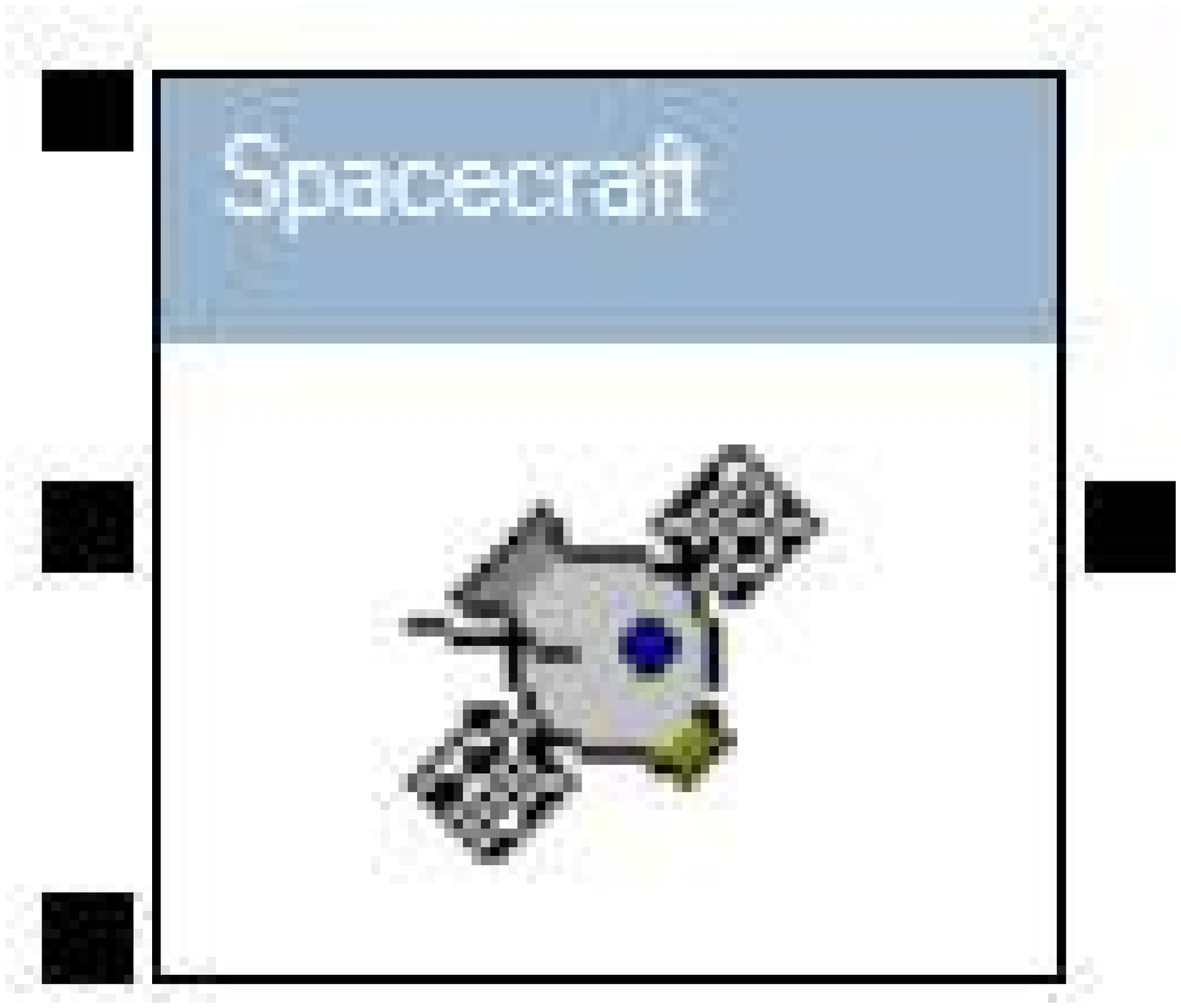}
\caption{Compact representation of spacecraft simulation model}\label{fig:spacecraft_2_elastic_3_gyro_simulation_compact}
\end{center}
\end{figure}

Now we have got simulation model of spacecraft. However this model is not facile for development of stabilization system. Control systems theory usually uses linearized models. Such model can be obtained by virtual vibration test. Note that presented on figure \ref{fig:spacecraft_2_elastic_3_gyro_simulation} is not facile. It contains a lot of big squares. Presented software supports compact representation of models. So we can replace model (figure \ref{fig:spacecraft_2_elastic_3_gyro_simulation}) by its compact representation (figure \ref{fig:spacecraft_2_elastic_3_gyro_simulation_compact})
During virtual vibration test or virtual spacecraft is being forced by momentum which satisfies PID control law \cite{pid_control} of momentum has been used:
\begin{equation}\nonumber
M_x=M_0(t)+K_1\omega + K_2\varphi + K_3\int\varphi d\varphi.
\end{equation}

Where $M_x$ is mechanical momentum,  is $X$ - projection of spacecraft angular velocity,  of spacecraft,   is rotation angle of spacecraft with respect to $X$ - axis. Parameters $K_1$, $K_2$, $K_3$ are constants.The test purpose is definition of spacecraft transformation function. Its definition can be obtained by response on harmonic input. The chirp input signal has been used:
\begin{equation}\nonumber
M_0(t)=C\sin(at+bt^2).
\end{equation}

\begin{figure}[htp]
\begin{center}
\hspace{-0.2cm}
\includegraphics[scale=0.2]{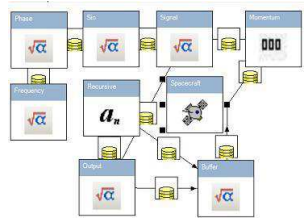}
\caption{Virtual vibration test scheme}\label{fig:virtual_vibration_test}
\end{center}
\end{figure}

\begin{figure}[htp]
\begin{center}
\hspace{-0.2cm}
\includegraphics[scale=0.3]{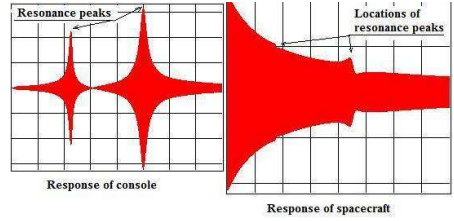}
\caption{Virtual vibration test scheme}\label{fig:virtual_vibration_test_response}
\end{center}
\end{figure}

Virtual vibration test scheme is presented on figure \ref{fig:virtual_vibration_test}
The scheme contains spacecraft's mechanical model and additional math which is necessary for virtual vibration test. In result of vibration test we have obtained response that is presented on figure \ref{fig:virtual_vibration_test_response}.
\subsection{Nonparametric identification}
\begin{figure}[htp]
\begin{center}
\hspace{-0.2cm}
\includegraphics[scale=0.4]{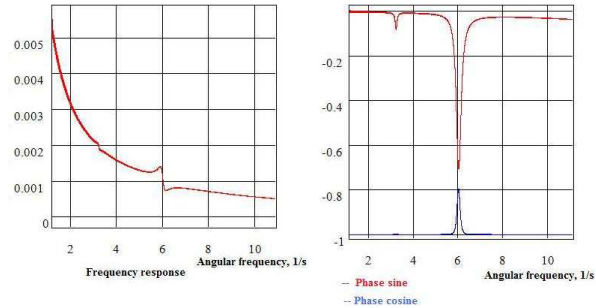}
\caption{Result of digital processing of virtual vibration test signals}\label{fig:nonparametric_identification}
\end{center}
\end{figure}

Nonparametric identification methods include definition of gain response and frequency response. Universal software can easy resolve these tasks. Full explanation reader can find article devoted processing of signals. Results of digital processing is presented in figure \ref{fig:nonparametric_identification}.
\subsection{Parametric identification}\label{sec:par_ident}
\begin{figure}[htp]
\begin{center}
\hspace{-0.2cm}
\includegraphics[scale=0.4]{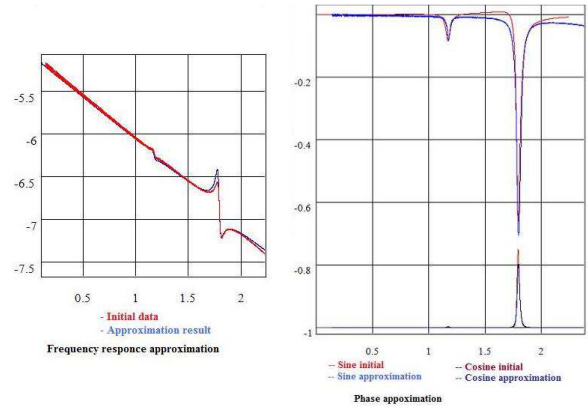}
\caption{Approximation results}\label{fig:parametric_identification}
\end{center}
\end{figure}

Parametric identification methods include definition of transfer functions. Control systems specialists use logarithmic scale for frequency response. This function provides clear picture of control object. So we transform functions of previous sections to logarithmic scale. Logarithmic transformations of gain response a frequency response (figure \ref{fig:nonparametric_identification}) are presented on figure \ref{fig:parametric_identification}.
The $Y$- axis of frequency response is also logarithmic. Control system specialist can define that such charts correspond to following transfer function:
\begin{equation}\nonumber
W(s)=\frac{k}{s}\frac{T_1s^2+T_2s+1}{T_3s^2+T_4s+1} \frac{T_5s^2+T_6s+1}{T_7s^2+T_8s+1}.
\end{equation}

\begin{figure}[htp]
\begin{center}
\hspace{-0.2cm}
\includegraphics[scale=0.4]{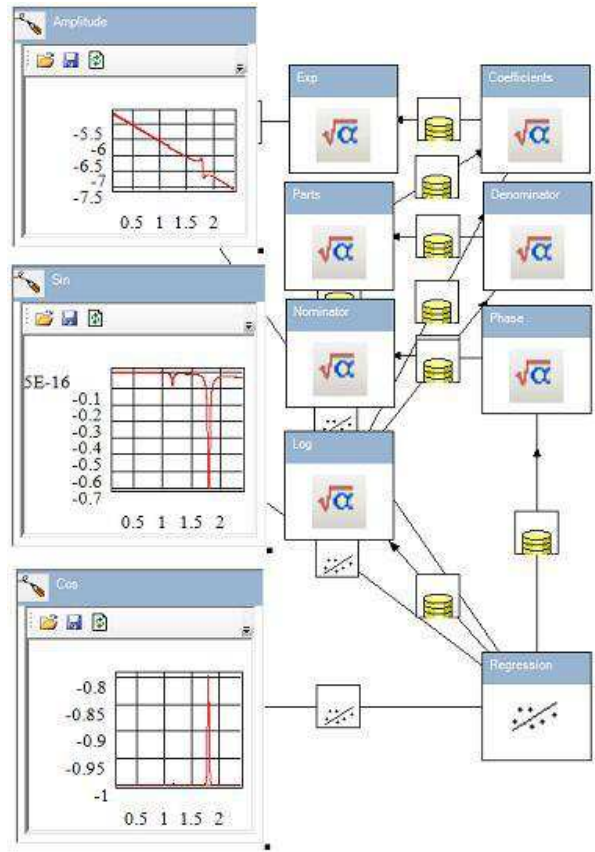}
\caption{Regression algorithm of parametric identification}\label{fig:parametric_identification_algorithm}
\end{center}
\end{figure}

But parameters $k$, $T_1$,...,$T_8$ are unknown. These parameters can be defined by nonlinear regression. Regression algorithm is presented in figure \ref{fig:parametric_identification_algorithm}.
Charts in the left part of figure \ref{fig:parametric_identification_algorithm} represents approximated functions (Frequency response, sine and cosine

 of phase). Other squares contain necessary math. Approximation result is presented in figure \ref{fig:parametric_identification}.
\subsection{Celestial navigation}
\begin{figure}[htp]
\begin{center}
\hspace{-0.2cm}
\includegraphics[scale=0.4]{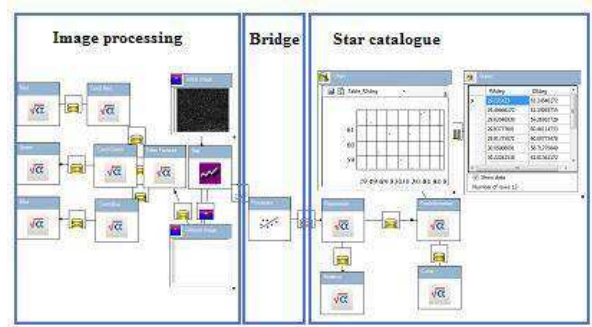}
\caption{Celestial navigation algorithm}\label{fig:celestial_navigation_algorithm}
\end{center}
\end{figure}

\begin{figure}[htp]
\begin{center}
\hspace{-0.2cm}
\includegraphics[scale=0.4]{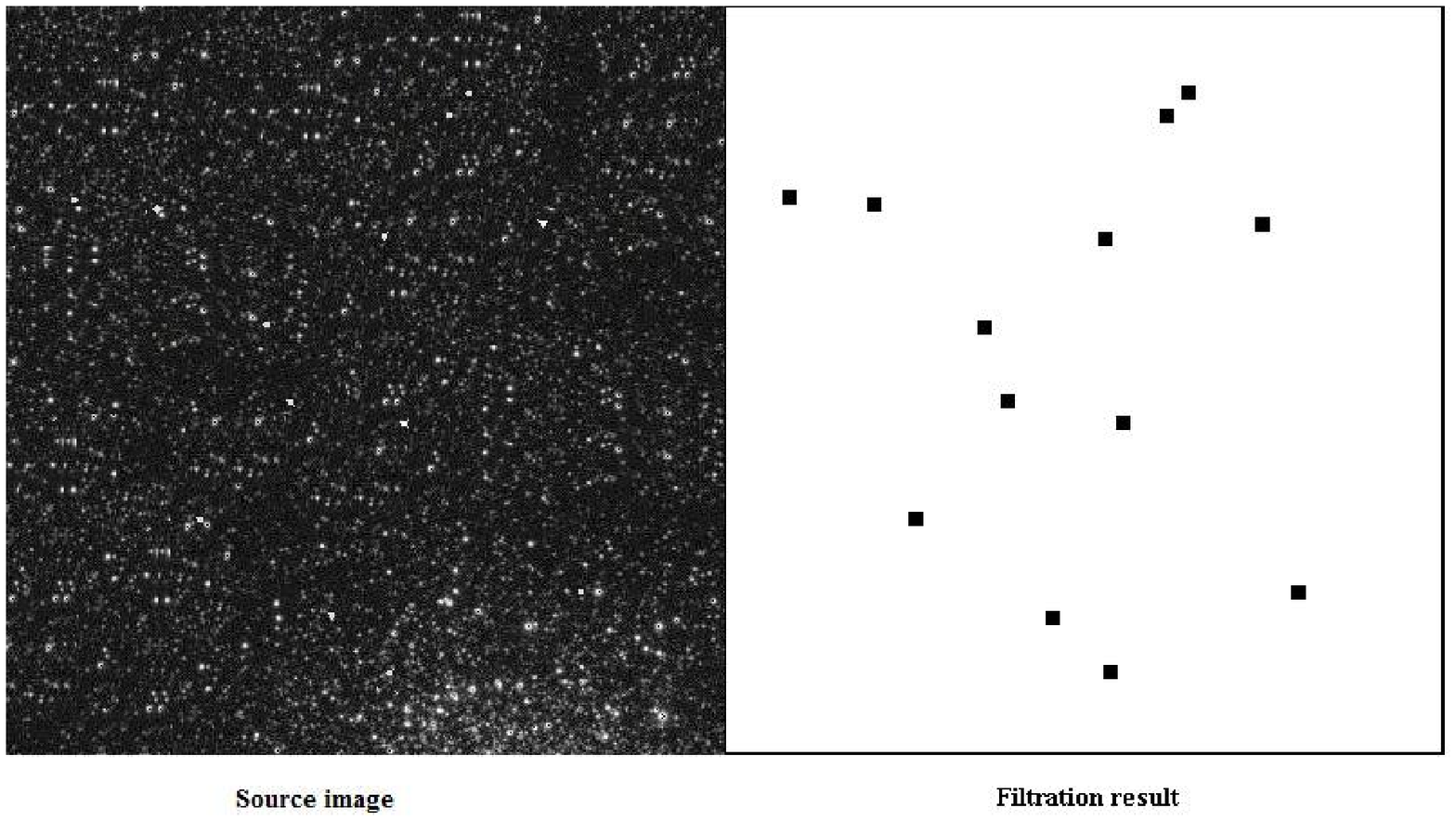}
\caption{Image filtration in celestial navigation algorithm}\label{fig:celestial_navigation_filtration_result}
\end{center}
\end{figure}

\begin{figure}[htp]
\begin{center}
\hspace{-0.2cm}
\includegraphics[scale=0.4]{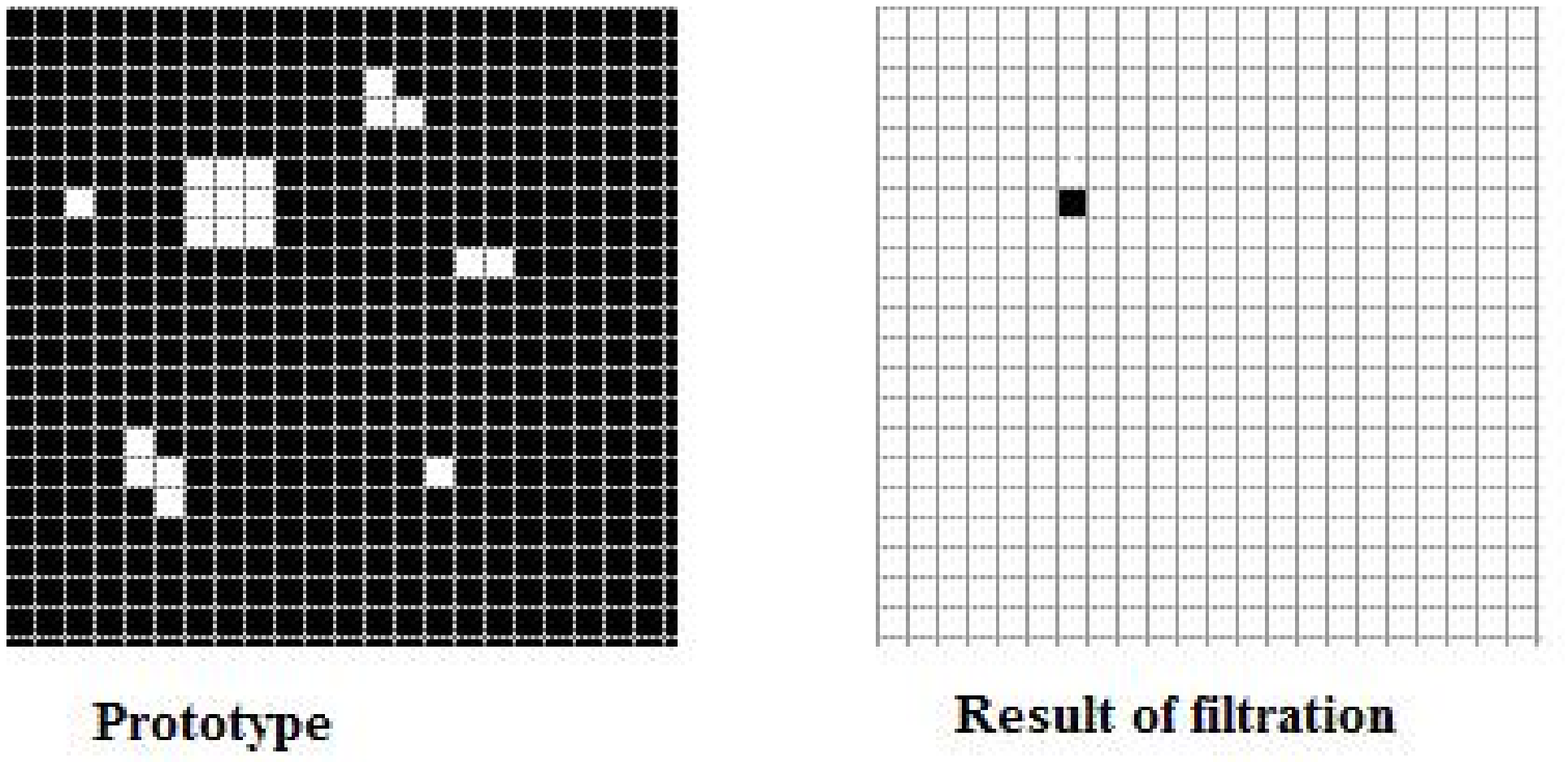}
\caption{Image filtration in celestial navigation algorithm}\label{fig:celestial_navigation_filtration}
\end{center}
\end{figure}

\begin{figure}[htp]
\begin{center}
\hspace{-0.2cm}
\includegraphics[scale=0.4]{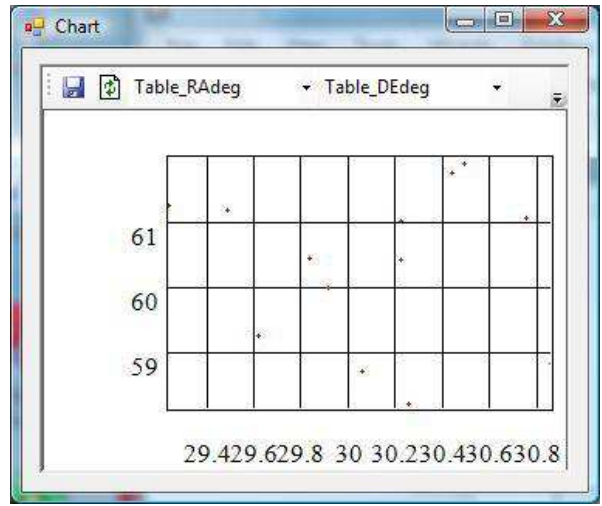}
\caption{Star catalogue query chart}\label{fig:celestial_navigation_sql_query_points}
\end{center}
\end{figure}

Any control system requires sensor. We use celestial navigation sensor for spacecraft control. The sensor enables to define orientation of spacecrafts. There are a lot of types of celestial navigation sensors. Here one of possible schemes is provided. Suppose that we have equipment that provides celestial images and star catalogues. Comparison of image and catalogue enable us to define orientation of equipment. So we can define orientation of spacecraft. Algorithm of this sensor is presented in figure \ref{fig:celestial_navigation_algorithm}.
First of all let us consider image processing. Any celestial image contains interfering information. We need filtration for its exclusion. Nonlocal digital image processing is being used for this purpose.
This scheme (figure \ref{fig:celestial_navigation_algorithm}) contains $\mathbf{Initial\ image}$ (Source image) obtained by equipment. Little squares in figure \ref{fig:celestial_navigation_algorithm} provides necessary math. It result we have $\mathbf{Filtered\ image}$ (Filtration result). Both images are presented in figure \ref{fig:celestial_navigation_filtration_result}.
Figure \ref{fig:celestial_navigation_filtration} explains image filtration algorithm.
If we have 9 closed white pixels then we replace it by one black pixel. Every other pixels are white. Result of filtration enables us to obtain $X$ and $Y$ coordinates of black pixels. Then this numbers will be compared with star catalogue. The $\mathbf{Stat}$ component extracts this numbers from $\mathbf{Filtered\ image}$.
Star catalogue is stored in database. Necessary information can be extracted by SQL query. Query statement is presented below:

\begin{lstlisting}[frame=tb]{somecode}


SELECT RAdeg, DEdeg FROM hip_main WHERE RAdeg > @RAMIN AND RAdeg < @RAMAX AND DEdeg >  @DEMIN AND DEDeg < @DEMAX AND BTmag > @BTMIN ORDER BY RAdeg

\end{lstlisting}
This statement has following meaning. First of all we consider limited area of sky. Declination and right ascension belong to small intervals. Secondly we consider only susch stars which magnitudes exceed defined constant (in this sample the constant is equal to 9). Query result provides following chart presented in figure \ref{fig:celestial_navigation_sql_query_points}.

We would like compare this chart with filtered image. This operation requires a set of math transformations. Essential feature of these transformations is Euclidean transformation:
\begin{equation}\nonumber
x'=(x+a)\cos\varphi+(y+b)\sin\varphi;
\end{equation}
\begin{equation}\nonumber
y'=-(x+a)\sin\varphi+(y+b)\cos\varphi;
\end{equation}
Parameters $a$, $b$, and $\varphi$  are unknown. Comparison of star catalogue and filtered image enable us to define these parameters. Using these parameters we can define orientation of spacecraft.
\subsection{Motion of mass center}
\begin{figure}[htp]
\begin{center}
\hspace{-0.2cm}
\includegraphics[scale=0.4]{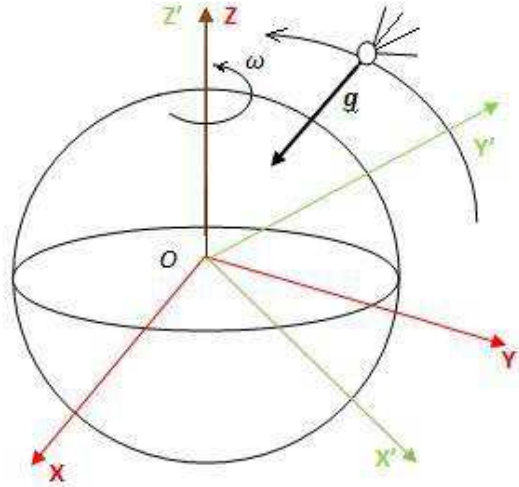}
\caption{Inertial reference geometry}\label{fig:spacecraft_inertial_picture}
\end{center}
\end{figure}

Since we consider magnetic method gyro momentum of compensation we should have model of Earth's magnetic field. But Earth's magnetic field induction depends near spacecraft depends on spacecraft position. So we need consider motion of spacecraft as a point.   Here I consider usage of inertial reference frame. It is more convenient for some tasks. Relation between these reference frames is shown in figure \ref{fig:spacecraft_inertial_picture}

\begin{figure}[htp]
\begin{center}
\hspace{-0.2cm}
\includegraphics[scale=0.4]{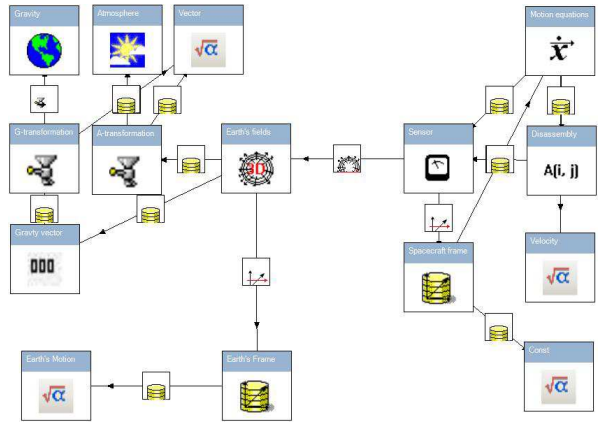}
\caption{Inertial reference frame geometry}\label{fig:spacecraft_inertial_simulation}
\end{center}
\end{figure}

The $OXYZ$ is inertial frame and $OX'Y'Z'$ is Greenwich one. Greenwich frame is rotated with respect to inertial one. We should transform acceleration vector $\mathbf{g}$ from Greenwich frame to inertial frame. Calculation of $\mathbf{g}$ in inertial reference frame has two features. Satellite coordinates with respect to $OXYZ$ are not equal to coordinates with respect to $OX'Y'Z'$. Moreover projections of $\mathbf{g}$ on $OXYZ$ axes of coordinates are different to projections on OX'Y'Z' axes of coordinates. Declarative approach enables us to resolve both problems at once. Usage of  covariant fields provides solution of both problems Simulation of linear satellite motion is presented in figure \ref{fig:spacecraft_inertial_simulation}.

Here $\mathbf{Earth's\ Frame}$ component represents Greenwich reference frame. The $\mathbf{Gravity}$  component is obtained from dynamically linked library. This component calculates Earth's gravity field/  Here this component is used as physical field (in Earth's fields). We can say the same about $\mathbf{Atmosphere}$ which calculates parameters of dynamical atmosphere model. Physical field is linked to Greenwich reference frame. Gravity field is marked as covariant. These circumstances provide  solution of both above problems. The Sensor is linked to Spacecraft frame. Its orientation coincides with orientation of inertial reference frame. Sensor results are used in Motion equations of spacecraft. Otherwise Motion equations results are used by Spacecraft frame. So we have math model of spacecraft linear motion.

\begin{figure}[htp]
\begin{center}
\hspace{-0.2cm}
\includegraphics[scale=0.1]{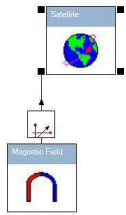}
\caption{Encapsulated of spacecraft motion model and magnetic field model}\label{fig:spacecraft_inertial_with_magnetic}
\end{center}
\end{figure}

The picture in figure \ref{fig:spacecraft_inertial_simulation} is not facile for further development since it contains a lot of squares. Some of squares should be encapsulated. Encapsulation means that we make invisible some elements in figure \ref{fig:spacecraft_inertial_simulation}
Now we would like add model of Earth's magnetic field. Result of this operation is presented in figure \ref{fig:spacecraft_inertial_with_magnetic}
We have full mechanical model. Next step is development of stabilization system. Development of stabilization system is based on results which was considered in \ref{sec:par_ident}. Physical principles of control system will be considered in \ref{fin_res}.
\subsection{Final result}\label{fin_res}
\begin{figure}[htp]
\begin{center}
\hspace{-0.2cm}
\includegraphics[scale=0.5]{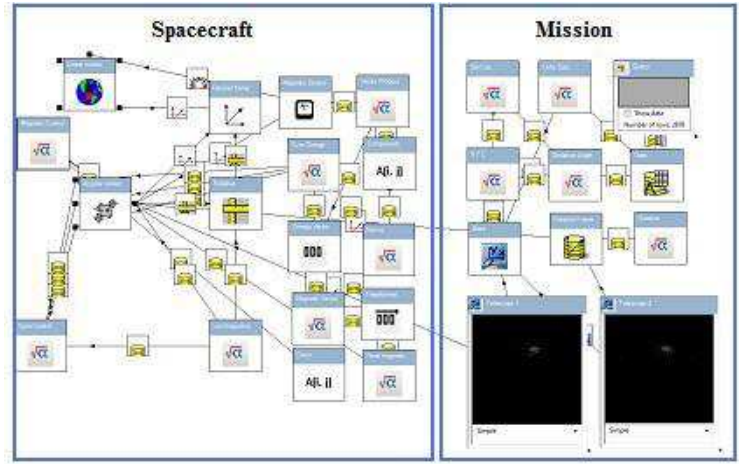}
\caption{Spacecraft and mission}\label{fig:spacecraft_and_mission}
\end{center}
\end{figure}

\begin{figure}[htp]
\begin{center}
\hspace{-0.2cm}
\includegraphics[scale=0.2]{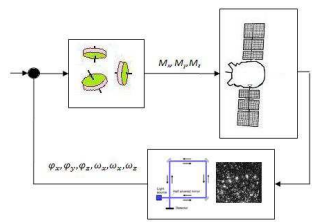}
\caption{High frequency loop}\label{fig:high_frequency_loop}
\end{center}
\end{figure}

\begin{figure}[htp]
\begin{center}
\hspace{-0.2cm}
\includegraphics[scale=0.2]{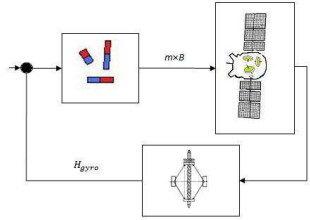}
\caption{Low frequency loop}\label{fig:low_frequency_loop}
\end{center}
\end{figure}

\begin{figure}[htp]
\begin{center}
\hspace{-0.2cm}
\includegraphics[scale=0.3]{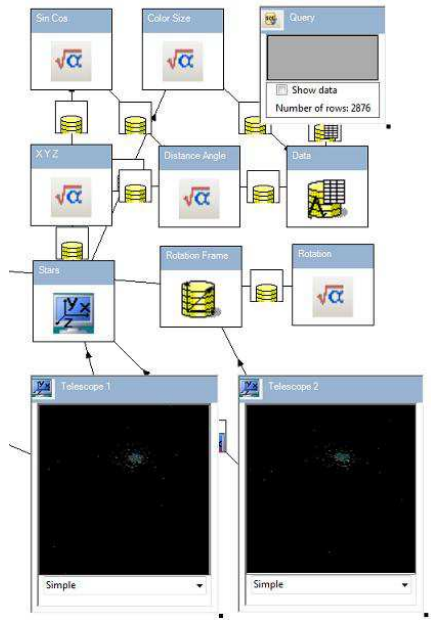}
\caption{Spacecraft mission}\label{fig:spacecraft_mission_pure}
\end{center}
\end{figure}

\begin{figure}[htp]
\begin{center}
\hspace{-0.2cm}
\includegraphics[scale=0.2]{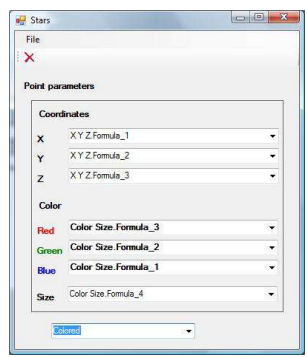}
\caption{Visualization component properties}\label{fig:properties_of_star_vizualization}
\end{center}
\end{figure}

Since this article is more illustration of usefulness of advantages of top-down paradigm we leave details of full construction. More details you can find in \url{http://www.codeproject.com/KB/architecture/grandiose2.aspx}. Figure \ref{fig:spacecraft_and_mission} shows final result.
In brief figure \ref{fig:spacecraft_and_mission} contains two main parts. Left part contains full motion model of spacecraft. Full motion model includes:
\newline
-mechanical model of spacecraft;
\newline
-model of Earth's gravity field;
\newline
-model of Earth's atmosphere;
\newline
-model of Earth's magnetic field;
\newline
-model of Celestial navigation;
\newline
-model of stabilization system of spacecraft.
\newline
\newline
Stabilization system could not use flywheels only (See \ref{sec:adv_mech}). Otherwise there is an obstacle for construction of stabilization system which uses electromagnets only. Electromagnetic momentum is always perpendicular to magnetic induction B. But stabilization requires all directions of control momentums. So space technology uses double-loop systems which use both flywheels and electromagnets. Let us consider both loops of this system. First loop (I name it high frequency loop) is presented in figure \ref{fig:high_frequency_loop}
We use celestial navigation  for definition of orientation and optical gyroscope  for definition of angular velocity. In result we have following parameters $\varphi_x$, $\varphi_y$, $\varphi_z$, $\omega_x$, $\omega_y$, $\omega_z$. First three parameters are angle deviations with respect to desired axes X, Y, Z. Following three parameters are angular velocities with respect to same axes. Control momentums are obtained by flywheels. We have identified spacecraft angular motion model in \ref{sec:par_ident}. In accordance to identified model we develop control law. I drop details here. Here I note that control law of high frequency loop is designed in compliance with control theory. Second loop scheme is presented in figure \ref{fig:low_frequency_loop}
This loop purpose is limitation of flywheels' angular velocities. Sensor of this loop are tachometers of flywheels. Necessary momentum is provided by electromagnets . It is possible to use different control laws for this loop. But main idea of these laws is single In this article I have used following control law. Let be $H_{gyro}$ is total angular momentum. Then momentum of electromagnets is opposite to $H_{gyro}$. But electromagnets are not always switched on. If $|H_{gyro}|$ is too small then electromagnets are switched of. Otherwise if angle between Earth's magnetic induction $H_{gyro}$ and $H_{gyro}$ is too small then electromagnets cannot provide substantionally large momentum that is opposite $H_{gyro}$. Therefore if angle between $H_{gyro}$ and $B$ is too small then electromagnets are also switched off. Since action of electromagnets is not continuous I name this loop low frequency loop. Here I explain how magnetic momentum reduces angular velocities of flywheels. Magnetic momentum causes deviation of spacecraft orientation. High frequency loop tries to eliminate this deviation by changing of angular velocity of flywheels. High frequency loop tries to eliminate this deviation by changing of angular velocity of flywheels. Since electromagnetic momentum is opposite to $H_{gyro}$ changing of angular velocities is reducing of their values.
\newline
Right part in figure \ref{fig:spacecraft_and_mission} presents spacecraft mission. The mission is astronomical observations. Pure mission is shown in figure \ref{fig:spacecraft_mission_pure}
Spacecraft contains two telescopes. $\mathbf{Telescope\ 1}$ is rigidly attached to spacecraft and do not use additional frame. $\mathbf{Telescope\ 2}$ is rotated relatively spacecraft. So $\mathbf{Telescope\ 2}$ is installed on $\mathbf{Rotation\ frame}$. Otherwise $\mathbf{Rotation\ frame}$ is installed on spacecraft body. Special component has been developed for indication of stars. Properties of the component are presented in figure \ref{fig:properties_of_star_vizualization}.
These properties have following meaning. Star coordinates $X$, $Y$, $Z$ are equal to $Formula\_1$, $Formula\_2$ $Formula\_3$ (of $\mathbf{X\ Y\ Z}$ component) respectively. Similarly color and size of star indication are defined. This sample requires Astronomy Express project and star catalogue (Hipparcos and Tycho).

 Both telescopes observes single collection of stars ($\mathbf{Stars}$ component). This collection is generated by following way. First of all Query component performs SQL query of star catalogue. Then other components perform necessary math transformations. Result of these transformations is used by Stars component.

\section{Conclusion}
Typical way of field of activity contains two stages. During first stage there is information acquisition and experience without understanding of fundamental principles. Many almost equivalent results are obtained independently by different researchers and/or developers. Then fundamental principles become well known and chaotic evolution changes to planned development. For example design patterns of GoF \cite{design_patterns} had being developed independently. After GoF these fundamental principles of software development became well known. I think that science and engineering software do not reach second stage yet in general. There are good fundamental principles of CAD, CAE etc. But they are not principles of all science and engineering software. I would like to show usefulness of such universal principles.

\end{document}